\documentstyle[revtex]{aps}
\begin{document}
%\draft

\begin{title}
  Topological Defects in Gravitational Theories \\
  with Non Linear Lagrangians
\end{title}
\author {J. Audretsch, A. Economou  and C.O. Lousto}
\begin{instit}
  Fakult\"at f\"ur Physik der
  Universit\"at Konstanz,
  Postfach 5560,
  D - 7750 Konstanz, Germany
\end{instit}

\begin{abstract}
 The gravitational field of monopoles, cosmic strings and domain walls is
 studied in the quadratic gravitational theory $R+\alpha R^2$ with $\alpha
 |R|\ll 1$, and is compared with the result in Einstein's theory.
 The metric  aquires modifications  which correspond to a short range
 `Newtonian' potential for gauge cosmic strings, gauge monopoles and domain
 walls and to a long range one for global monopoles and global cosmic
 strings. In this theory the corrections turn out to be attractive for all
 the defects. We explain, however, that the sign of these corrections
 in general depends on the particular higher order
 derivative theory and topological defect under consideration. The possible
 relevance of our results to the study of the evolution of topological
 defects in the early universe is pointed out.

\end{abstract}


%\narrowtext

\section{Introduction}

   After the paradigm of the Hilbert's Lagrangian formulation of
 Einstein's theory of gravity it was clear how one
 could consistently formulate other,
 higher derivative gravitational theories (that is theories in
 which the field equations have higher than second metric derivatives).
 And such theories where indeed proposed and used as alternatives to
 Einstein's theory in attempts to unify other fields with
 gravity  \cite{Weyl}  and to remedy some of its seemingly undesirable
 consequences  as, for example, at the classical level, the unavoidance
 of cosmological singularities  \cite{Kerner} and, at the quantum level,
 the non renormalizability of the quantized version of general
 relativity \cite{Stelle}.

 One of the main motivations for studying higher derivative theories
 comes from the  semiclassical general relativity.
 There,  it seems to be  a matter of self-consistency to consider higher
 derivative terms in the gravitational Lagrangian  since such terms
 arise generically in one-loop calculations \cite{Birrell}.
 Certainly this notion of self-consistency is a delicate issue and, as
 Simon has recently suggested \cite{Simon},  it needs to be reconsidered
 if one wants to avoid undesirable semiclassical predictions such  as
 unstable Minkowski spacetime.
 Another recent motivation for considering higher derivative
 gravitational theories is that such theories have arisen as low energy
 limit of several superstring theories \cite{Superstring}.

 Higher derivative theories are of interest to cosmology mainly because,
 even vacuum theories admit cosmological models which give rise to the,
 so called, Starobinsky inflation \cite{Starobinsky} (see however
 Ref. \cite{Simon} for a  critisism on its consistency in the semiclassical
 limit), without fine tuning of the initial conditions \cite{Feldman}.

 In this paper we want to look at another topic of cosmological
 relevance namely, the effects of higher derivative theories on the
 gravitational field of topological defects as monopoles, cosmic strings
 and domain walls.
 These are objects that may have formed during phase transitions in the
 cooling down of the Early-Universe and    may have played a key role in
 the formation of the large scale structure of the Universe mainly
 through their gravitational interactions \cite{Vilenkin85,Brandenberger}.
 Since their main interaction is gravitational, it is important to have
 an idea of what modifications one should expect in their gravitational
 field when the relevant gravitational theory has higher derivatives.
 Some work has recently been done in this direction, but only for gauge
 cosmic strings \cite{Linet}.
 This work shows that  in the weak  field limit only short range
 corrections to the Einstein theory arise which
 are associated with the presence of additional massive fields in the
 spectrum of higher derivative theories.
 However, this is not expected to be in general true, especially for
 global topological defects which are extended field configurations and
 not localized as the gauge cosmic strings.

 In this work we have in mind  theories that can be separated in a part
 ${\cal L}_G$ for the gravitational field $g_{\mu\nu}$ constructed with
 geometrical scalars of the Ricci tensor $R_{ab}$, and another  part
 ${\cal L}_M$ containing  matter fields with standard coupling to the
 gravitational field $g_{\mu\nu}$
\begin{equation}
 {\cal L} = {1\over 2\kappa}{\cal L}_G(R_{ab}) + {\cal L}_M(g_{ab}).
\label{FullLagrangian}\end{equation}
 Hereafter $\kappa:=8\pi G$ where $G$ is the gravitational constant.
 For theories of this type it has been  noted that  they can be recasted
 into an equivalent theory  of Einstein gravity  interacting with
 additional matter fields \cite{Whitt,Magnano,Jakubiec}. However, as it
 was stressed by Brans \cite{Brans} and we shall explain in the next
 section, this  equivalence is in general only at a mathematical level
 and not at a physical one. Nevertheless, based on such an equivalent
 system, Whitt \cite{Whitt} was able to show that the black holes of
 general relativity are the only black hole solutions of  $R+R^2$
 theories (no hair theorem).

 For the  discussion of this paper we will deal  with theories that have
 as gravitational part  the following, often appearing in the
 literature, Lagrangians
\begin{equation}
    {\cal L}_G = \sqrt{-g}
    ( R + \alpha R^2 + \beta R_{\mu\nu} R^{\mu\nu} )
\label{QuadraticLagrangian}\end{equation}
 and
\begin{equation}
    {\cal L}_G = \sqrt{-g} F(R) ,
\label{ScalarQuadraticLagrangian}\end{equation}
 where  $\alpha, \beta$ are some coupling constants,
 $g:={\rm det}g_{\mu\nu}$, and $R=g^{\mu\nu}R_{\mu\nu}$. Finally the $F$
 in Eq.\ (\ref{ScalarQuadraticLagrangian}) is in principle an arbitrary
 function of the curvature scalar $R$. However, later on we will
 take $F$ to differ only slightly from the Einstein value $R$, that is
 $F=R+\alpha R^2$ with $\alpha |R|\ll 1$.
 See Ref. \cite{Whitt} for a treatment of the  $F=R+\alpha R^2$ theory in
 vacuum and the Ref. \cite{Maeda} together with references therein for
 generalizations to arbitrary $F(R)$ in the presence of particular forms
 of matter.

 The structure of the paper is as follows. Section II contains a brief
 review of higher derivarive theories to the extent needed in
 this paper. The field equations for the
 theories in (\ref{FullLagrangian}),(\ref{QuadraticLagrangian}) and
 (\ref{FullLagrangian}),(\ref{ScalarQuadraticLagrangian}),
 are written down, and
 their spectrum is explained. With  the procedure that enables the
 recasting of these field equations  into  Einstein type ones we obtain
 the basic result that is used  in the Sec. III for the comparison of
 the gravitational field of global monopoles, cosmic strings and domain
 walls in Enstein's theory, and  in the quadratic  $R+\alpha R^2$
 theory with $\alpha |R|\ll 1$. Section III also contains at the
 beginning a brief introduction to the topological defects.
 Finally in Sec. IV we conclude with a brief summary and comments.

 Throughout this paper we use the conventions $\hbar=c=1$, metric
 signature $(-+++$), Riemann tensor $R{}^a{}_{bcd}:=- \partial{}_d
 \Gamma{}^a{}_{bc}+ \dots$, and Ricci tensor
 $R{}_{ab}:=R{}^c{}_{acb}$.

\section{Theories with higher derivatives}

\subsection{Field equations}

 We shall give now the gravitational field equations for the higher
 derivative theories given by   the
 Eqs.\ (\ref{FullLagrangian}), (\ref{QuadraticLagrangian})
 and  Eqs.\ (\ref{FullLagrangian}), (\ref{ScalarQuadraticLagrangian}).
 The field equations  for $g_{\mu\nu}$ are obtained by varying the
 action corresponding to Eq.\ (\ref {FullLagrangian})
 with respect to $g^{\mu\nu}$
 and contain derivatives of the metric up to the fourth order. For the
 case of ${\cal L}_G$ of  Eq.\ (\ref{QuadraticLagrangian}) they read
\begin{eqnarray}
     (1 &+&2\alpha R)(R_{\mu\nu} - {1\over 2}g_{\mu\nu}R) +
          {\alpha \over 2}R^2  g_{\mu\nu}    \cr
     &+&(2\alpha + {\beta\over 2})g_{\mu\nu} R_{;p}{}^{;p}
         -(2\alpha+\beta) R_{;\mu\nu}         \cr
   &+&\beta  R_{\mu\nu;p}{}^{;p} -
   {\beta\over 2} R_{pq}R^{pq}g_{\mu\nu} +
  2\beta R_{pq}R{}_\mu{}^p{}_\nu{}^q  \cr
  && \qquad\qquad ={-2\kappa\over\sqrt{-g}}
         {\delta  S_M \over \delta (g^{\mu\nu})} :=
  \kappa T^{(M)}_{\mu\nu} .
\label{FieldEqs}\end{eqnarray}
 Notice that the trace of this equation is an inhomogeneous massive
 Klein-Gordon equation for the curvature scalar $R$
\begin{equation}
     (6\alpha+2\beta)R{}_{;p}{}^{;p} - R= \kappa T^{(M)}.
\label{TraceOfFieldEqs}\end{equation}

 Finally, the field equations  for the theory
 (\ref{FullLagrangian}) and (\ref{ScalarQuadraticLagrangian}) can be
 written as
\FL
\begin{equation}
    F^\prime G_{\mu\nu} = \kappa T^{(M)}_{\mu\nu} + {1\over 2}
  g_{\mu\nu}(F-F^\prime R- 2 F^\prime_{;p}{}^{;p}) +
  F^\prime_{;\mu\nu} ,
\label{ScalarFieldEqs}\end{equation}
 where $F^\prime= \partial F/  \partial R$ and $G_{\mu\nu}= R_{\mu\nu}
 -{1\over 2}g_{\mu\nu} R$ is the Einstein tensor. The trace of this
 equation  is
\begin{equation}
 3 F^\prime_{;p}{}^{p} + F^\prime R-2F = \kappa T^{(M)}.
\label{TraceOfScalarFieldEqs}\end{equation}

\subsection{Spectrum of quadratic theories - \\ Weak gravitational limit}

 We would like  to stress here the fact that quadratic theories  do not
 contain only the usual massless (long-range) spin-2 graviton field but
 also, in general,  two massive (short-range) fields with spin-0 and
 spin-2.

 This spectrum  can be  easily recognized in the case of ${\cal L}_G$ of
 Eq.\ (\ref{QuadraticLagrangian}) when one writes the field equations  in
 the linearized weak field  limit using a convenient gauge (coordinate
 system). Indeed following Teyssandier \cite{Teyssandier} we have that
 $g_{\mu\nu}$ can be decomposed in the weak gravitational limit (where
 $g_{\mu\nu}=\eta_{\mu\nu} + h_{\mu\nu}$ with $|h_{\mu\nu}| <<1$) as
\begin{equation}
   g_{\mu\nu}= \eta_{\mu\nu} + h^{\rm (E)}_{\mu\nu} +
             \chi\eta_{\mu\nu} +\psi_{\mu\nu} ,
\label{LinearizedEqsa}\end{equation}
 with the field equations
\FL
\begin{eqnarray}
          &\Box & h^{\rm (E)}_{\mu\nu} =
      -2\kappa( T_{\mu\nu}-{1\over 2}T \eta_{\mu\nu} ) ,
      \cr
        (&\Box& - m_0^2 )\chi = -{1\over 3} \kappa T ,
       \quad m_0^{-2}:=6\alpha+2\beta ,
      \cr
       (&\Box & - m_1^2 )\psi_{\mu\nu} =
                2\kappa( T_{\mu\nu}-{1\over 3}T \eta_{\mu\nu} ),
         \quad m_1^{-2}:=-\beta ,
\label{LinearizedEqsb}\end{eqnarray}
 and the gauge conditions
\begin{eqnarray}
    \partial^a( h^{\rm (E)}_{\mu a} - {1\over 2}
    h^{{\rm (E)}\lambda}_\lambda \eta_{\mu a})=0 ,
      \cr
     ( \psi^{ab} - \psi^\lambda_\lambda \eta^{ab})_{,ab} = 0 .
\label{LinearizedEqsc}\end{eqnarray}
 Here  indices are raised and lowered with the Minkowski metric tensor
 and the operator $\Box$ is  the Minkowskian one.  One recognizes in
 Eqs.\ (\ref{LinearizedEqsa})-(\ref{LinearizedEqsc})
 the usual Einstein contribution
 $h^{\rm (E)}_{\mu\nu}$, that is the graviton field which has 2 degrees
 of freedom.
 Then, a scalar field $\chi$ with mass $m_0$, which  obviously has one
 degree of freedom and appears as an overall conformal factor (in the
 considered approximation). Finally the massive tensorial field
 $\psi_{\mu\nu}$ with mass $m_1$ which turns out to have five degrees of
 freedom (note that in contrast to $h^{\rm (E)}_{\mu\nu}$ its components
 satisfy only one gauge condition) and thus it posseses the stucture of
 a massive spin-2 field.
 In order to keep the  ``mass'' parameters $m_0, m_1$  real  we shall
 demand the {\sl no-tachyon} constraint
\begin{equation}
 3\alpha + \beta \geq 0, \qquad \beta\leq 0.
\label{NoTachyon}\end{equation}

 We leave for the next subsection the case of the Lagrangian
 in Eq.\ (\ref{ScalarQuadraticLagrangian})
 where we will go beyond the weak
 gravitational limit and we will see that the spectrum of this theory
 consists of a graviton and a massive interacting scalar  field.

\subsection{Reformulation of quadratic theories}

 Interestingly enough, besides $g_{\mu\nu}$ there is an alternative
 canditate for  the metric field of the
 spacetime \cite{Magnano,Jakubiec}, namely the $\gamma_{\mu\nu}$ which
 is the inverse of $\gamma^{\mu\nu}$ where
\begin{equation}
    \sqrt{-\gamma}\gamma^{\mu\nu}:=
 { \partial {\cal L}_G \over  \partial  R_{\mu\nu}} ,
\label{NewMetricDefinition}\end{equation}
 and $\gamma:={\rm det}\gamma_{\mu\nu}$.

 In particular for the ${\cal L}_G$ of Eq.\ (\ref{QuadraticLagrangian})
 we have
\FL
\begin{equation}  \sqrt{-\gamma}\gamma^{\mu\nu}:=
            \sqrt{-g} \Bigl[ (1+2\alpha R)g^{\mu\nu} +
             2\beta R_{\alpha\beta}g^{\alpha\mu}g^{\beta\nu}\Bigr ] .
\label{NewMetric}\end{equation}
 Expressing the theory in terms of $\gamma_{\mu\nu}$ via a Legendre
 transformation, one can reduce the order of the derivatives that appear
 in the field equations  from fourth to second. But what is also
 important is, that the resulting theory takes the form of Einstein
 gravity for the metric $\gamma^{\mu\nu}$ plus some additional massive
 fields interpreted as matter fields.
 Thus the equation (\ref{NewMetric}) can be considered as a non linear
 decomposition of $g^{\mu\nu}$ in the physical spectum of the full
 theory:  a spin-2 massless field $\gamma^{\mu\nu}$, a scalar field
 that appears as a conformal factor and is a linear function of $R$,
 and finally, a tensor
 field related to $R^{ab}$ with 5 degrees of freedom ($R^{ab}$ is
 symmetric satisfying the 4 contracted Bianchi equations and its trace
 is essentially the previously mentioned scalar field).

 We can be more explicit in the  interesting case  of the theory in
 Eq.\ (\ref{FullLagrangian}) where ${\cal L}_G$ is given by
 Eq.\ (\ref{ScalarQuadraticLagrangian}).
 Here, only an additional scalar field appears since $\gamma^{\mu\nu}$
 and $g^{\mu\nu}$ are conformally related. Indeed
 Eq.\ (\ref{NewMetricDefinition}) gives
\begin{equation}  \sqrt{-\gamma}\gamma^{\mu\nu}:=
               \sqrt{-g} F^\prime g^{\mu\nu}  ,
\label{NewConformalMetric}\end{equation}
 which implies that
\begin{equation} \gamma_{\mu\nu}= F^\prime{}^{-1} g_{\mu\nu} .
\label{ConformalMetric}\end{equation}
 Defining a scalar field $\psi$ (not to be confused with the tensorial
 field $\psi_{\mu\nu}$ of Eqs.\ (\ref{LinearizedEqsa}),
 (\ref{LinearizedEqsb}) via
\begin{equation}
 F^\prime= \exp\left ( \sqrt{2\kappa\over 3}\psi \right ) ,
\label{psiField}\end{equation}
 the field equations (\ref{ScalarFieldEqs}) are written in the system
 $\gamma_{\mu\nu}$ as
\begin{equation}
 \widehat G_{\mu\nu} = {\kappa\over F^\prime} T^{(M)}_{\mu\nu}(g_{ab}) +
         \kappa T_{\mu\nu}^{(\psi)}
\label{TransformedFieldEqs}\end{equation}
 where, hereafter, hats denote quantities with respect to the metric
 $\gamma_{\mu\nu}$ and
\begin{equation}
    T_{\mu\nu}^{(\psi)}=  \widehat\nabla_\mu\psi \widehat\nabla_\nu\psi
     -{1\over 2} \gamma_{\mu\nu}\widehat\nabla_\lambda\psi
       \widehat\nabla^\lambda\psi -{1\over 2} \gamma_{\mu\nu} U(\psi).
\label{psiSEMtensor}\end{equation}
 The potential $U(\psi)$ is given by
\begin{equation}
 U(\psi)={1\over 2\kappa} F^\prime{}^{-2}(RF^\prime-F) ,
\label{Potential}\end{equation}
 which can be written as a function of $\psi$ alone in regions where
 Eq.\ (\ref{psiField}) is invertible.
 Finally  the scalar field $\psi$ satisfies the equation
\FL
\begin{equation}
 \widehat{\Box\vbox{\vskip 6pt}}\psi = \biggl (
 {\kappa\over 6}\biggr)^{1/2}
     F^\prime{}^{-2} \biggl [ {2F-RF^\prime\over \kappa}
      + g^{\mu\nu} T^{(M)}_{\mu\nu}(g_{ab})  \biggr ],
\label{psiFieldEq}\end{equation}
 which can be checked to be equivalent to the trace
 (\ref{TraceOfScalarFieldEqs}) of the initial field equations
 (\ref{ScalarFieldEqs}).

 These field equations (\ref{TransformedFieldEqs}) and (\ref{psiFieldEq})
 follow from the Lagrangian
\begin{eqnarray}
 {\cal L}^\prime = &&{1\over 2\kappa}\sqrt{-\gamma}
  \widehat{R}(\gamma_{ab})  \cr
  &+&\sqrt{-\gamma}[-{1\over 2}
  \gamma^{\mu\nu}\psi_{,\mu}\psi_{,\nu} -U(\psi)] +
  {\cal L}_M(g_{ab}) ,
\label{NewLagrangian}\end{eqnarray}
 which shows that  the quadratic theory (\ref{FullLagrangian}) is, loosely
 speaking, equivalent to ``Einstein's'' gravitational theory for the
 metric $\gamma_{ab}$ plus an interacting massive scalar field $\psi$
 plus the ``peculiar'' (not anymore usual) matter fields of
 ${\cal L}_M(g_{ab})$. They are indeed peculiar if $\gamma_{ab}$ is
 considered as the metric of spacetime since then  the dependence of
 ${\cal L}_M$ on $g_{ab}$ implies, via Eq.\ (\ref{ConformalMetric}), a non
 standard interaction of the metric field $\gamma_{ab}$ and of the field
 $\psi$ with the matter fields of ${\cal L}_M$. More on this issue of
 equivalence will be said at the end of this section.

\noindent\underbar {Case: $F(R) \approx R +\alpha R^2 $ with
                    $\alpha |R|\ll  1$.}

 \noindent We will now consider the interesting case where $F(R)$ can be
 expanded as a Taylor series around $R=0$ and deviates only slightly
 from Einstein's theory
\FL
\begin{equation}
   F(r)= R + \alpha R^2 + O(R^3),\quad \alpha |R|\ll 1,
   \quad
   \alpha:= %{1\over 2}{ \partial^2 F\over \partial R^2}\biggr \vert_{R=0}.
{F^{\prime\prime}\over 2}\bigr \vert_{R=0}.
\label{ApproxF}\end{equation}
 Assuming that the $O(R^3)$ terms can be ignored in this expression then
 the field equations (\ref{TransformedFieldEqs})
 and (\ref{psiFieldEq}) simplify
 considerably. Indeed, in this approximation  Eqs.\ (\ref{psiField}),
 (\ref{ApproxF}) imply
\begin{equation}
   \psi \approx \left (6\over \kappa \right)^{1/2}\alpha  R,
\label{psiApprox}\end{equation}
 while the leading term in the potential $U(\psi)$
 of Eq.\ (\ref{Potential})
 is,  assuming the no-tachyon constraint $\alpha>0$, a mass term
\begin{equation}
 U(\psi) = {1\over 2}m_0^2\psi^2 + O(\psi^3),
 \qquad    m_0^2 = {1\over 6\alpha}.
\label{PotentialApprox}\end{equation}
  The metric $\gamma_{\mu\nu}$ and the field $\psi$ are obtained from the
 field equations (\ref{TransformedFieldEqs}) and (\ref{psiFieldEq}) which,
 to lowest order in the approximation (\ref{ApproxF}), read
\begin{eqnarray}
        \widehat G_{\mu\nu}&\approx & \kappa T^{(M)}_{\mu\nu}(\gamma_{ab}),
    \cr
         (\widehat{\Box\vbox{\vskip 6pt}} - m_0^2)\psi &\approx &
          \left ( {\kappa\over 6}\right)^{1/2} T^{(M)}(\gamma_{ab}).
\label{ApproxFieldEqs}\end{eqnarray}
 Making use of Eq.\ (\ref{ConformalMetric}) one can finally obtain the
 metric $g_{\mu\nu}$.
 Notice that, to the considered approximation,  it does not matter which
 metric we actually use in  $T^{(M)}_{\mu\nu}$. It is more convenient,
 however, from the technical point of view to use $\gamma_{\mu\nu}$.

 Concluding we arrive at the following result:

 \noindent
 For a given matter source $T^{(M)}_{\mu\nu}(g_{ab})$, the metric
 $g_{\mu\nu}$ in the quadratic theory
 (\ref{FullLagrangian}) and (\ref{ScalarQuadraticLagrangian})
 with $F(R)\approx R +\alpha R^2$ and $\alpha |R|\ll 1$ is given by
\begin{equation}
 g_{\mu\nu}=[1+\chi]\gamma_{\mu\nu} ,
\label{QuadraticFromEinsteinMetric}\end{equation}
 where $\gamma_{\mu\nu}$ is the metric in the Einstein's theory with
 source $T^{(M)}_{\mu\nu}(\gamma_{ab})$ while the field $\chi$
 satisfies the equation
\begin{equation}
 ({\Box} - m_0^2)\chi =
    -{\kappa\over 3} T^{(M)}(\gamma_{ab}), \qquad m_0^2:={1\over 6\alpha}
\label{chiFieldEq}\end{equation}
 with the ${\Box}$ operator taken with respect to the $\gamma_{ab}$
 metric.

 This result follows directly from Eqs.\ (\ref{ApproxFieldEqs}),
 (\ref{ConformalMetric})
 using the variable $\chi$ related to the field $\psi$ via
 $\chi:=-2[\kappa/6]^{1/2}\psi$.
 Note that from (\ref{psiApprox}) follows $\alpha R=-\chi/2$ and therefore
 the condition $\alpha |R| \ll 1$ for our approximation is equivalent to
 $|\chi/2|\ll 1$.
 Finally, let us notice that  in the weak gravitational limit  the
 Eqs.\ (\ref{QuadraticFromEinsteinMetric}) and (\ref{chiFieldEq})
 are consistent with the
 $\beta = 0$ limit of the linearized equations (\ref{LinearizedEqsa}) and
 (\ref{LinearizedEqsb}).

\subsection{Some remarks}

 Based on the decomposition (\ref{NewMetricDefinition}) several
 authors \cite{Magnano,Ferraris88,Jakubiec,Schmidt,Maeda} have dealt
 with the question of whether quadratic theories are equivalent to
 Einstein's theory plus some additional fields. It seems that this may
 well be true for vacuum  theories. However, as was pointed out by
 Brans \cite{Brans}, (see also Refs.\cite{Sokolowski,Ferraris90}), a
 subtlety appears in the case where usual matter is present.
 The problem is  that the equivalence principle, a basic guide that one
 may use in constructing theories coupled to gravity and in particular
 to Einstein's theory, cannot be valid in both the original and the
 reformulated theories.
 If it is valid in the original theory, then a test matter  in
 ${\cal L}_M$ of the Lagrangian (\ref{FullLagrangian}) will follow
 geodesics of the spacetime with metric $g_{\mu\nu}$ but,
 in general, will fail
 to do the same in the spacetime with the metric $\gamma_{\mu\nu}$. In
 this sense we are not entitled to consider the reformulated theory as
 Einstein's theory in the presence of some interacting fields.
 In the case that one is philosophically inclined to consider the
 $\gamma_{\mu\nu}$ as the physical metric, while the $g_{\mu\nu}$ as
 some sort of unifying field, then the equivalence principle should be
 implemented in the matter part ${\cal L}_M$ of
 Eq.\ (\ref{FullLagrangian}) using
 the metric $\gamma_{\mu\nu}$ in the place of  $g_{\mu\nu}$. Of course
 then, this ${\cal L}_M$ will be non standard with respect to
 $g_{\mu\nu}$.

 Whether or not nature chooses  to couple usual matter universally only
 to a spin-2 field (as the $\gamma_{\mu\nu}$) and not to a more
 composite one (as the $g_{\mu\nu}$)  is far from being experimentally
 testable. Trying to find an answer one may, however, employ some
 criteria of principle, as positivity of energy \cite{Sokolowski}. In
 any case, the use of new variables, as those of Eq.\ (\ref{NewMetric}) and
 Eq.\ (\ref{ConformalMetric}), which  turn out to simplify technically a
 physical problem, is undoubtfully very useful even if it is not clear
 whether one can attribute to these variables a foundamental character.

\section{Topological defects \\ in higher derivative theories}

Cosmological defects are formed during phase transitions in the
 evolving Early Universe whenever the symmetry group $G$ of the relevant
 field theory breaks down to a subgroup $H$ so that the vacuum manifold
 $M=G/H$ has some non trivial homotopy group \cite{Kibble}.
 Such a symmetry breakdown  at an energy scale $\eta$ can be realized,
 e.g., with an $n$-component scalar field $\phi^{(i)}$ having a
 Mexican-hat type of potential
\begin{equation}
 V(\phi) = -{\lambda\over 4}(\sum_{i=1}^n\phi^{(i)}\phi^{(i)}- \eta^2)^2 .
\label{MexicanHatPot}\end{equation}
 The  homotopic structure of the vacuum manifold depends on the number
 $n$ of  components of the scalar field and, thus, we may have the
 formation of domain walls for $n=1$, cosmic strings for $n=2$,
 monopoles for $n=3$.
 These defects are  respectively  surface-, line-,  and point-like
 configurations. Sufficiently away from these configurations, at
 distances  $d\gg \delta$, the scalar field $\phi^{(i)}$ approaches
 quickly its vacuum value $\sum_i\phi^{(i)}\phi^{(i)} \approx \eta^2$.
 Here $\delta$ is the width of the core of these defects, of the order
 of $m_{\phi}^{-1}$ where $m_{\phi}=\eta\sqrt{\lambda}$ is the mass of
 the scalar field's massive mode. Typically, for symmetry breaking at
 grand unification scale, $\delta\approx 10^{-30}{\rm cm}$ and
 $\kappa\eta^2\approx 10^{-6}$.

 Depending on whether the symmetry that breaks down is a gauge (local)
 or a global one we have respectively the formation of {\sl gauge} or
 {\sl global} topological defects.
 In the case of gauge symmetry there exists a well defined core, with
 width $\delta$,  where most of the energy of the topological defect
 configuration is localized.
 On the other hand, for global topological defects the components of the
 respective stress-energy-momentum tensor have, outside the ``core'', a
 relatively slow fall off due to the gradients of the Goldstone modes of
 the scalar field $\phi^{(i)}$. Thus, global defects are extended
 configurations. The reason for this difference between gauge and global
 defects is that in the case of gauge symmetry the presence of gauge
 fields can compensate the gradients of the scalar field.
 Finally, in the case of discrete symmetry breaking, which gives rise to
 domain walls, there are no Goldstone modes and thus domain walls are
 localized configurations.

 Based on the above properties, we will make in what follows the
 following approximations:
 \begin{itemize}
\item[(i)] Gauge topological defects and domain walls will be considered
    in the  zero core-thickness approximation
    and thus their stress-energy-momentum tensors will have components
    with appropriate Dirac $\delta$-fuctions.
\item[(ii)] For global defects, we will make the $\sigma$-model
    approximation
    where the scalar field is fixed to its asymptotic vacuum value
    everywhere outside the defect. This  is a sensible approximation at
    distances from the defect sufficiently larger than the ``core''
    width $\delta$.
 \end{itemize}

 In the following subsections we will obtain the gravitational field of
 cosmic strings, monopoles and domain walls in the quadratic theory $R+
 \alpha R^2$ with $\alpha |R| \ll 1$. For this we will make use of the
 result of the previous section (see Eqs.
 (\ref{QuadraticFromEinsteinMetric}), (\ref{chiFieldEq})), stating that
 the metric in the quadratic theory, $ds^2_{\rm (Q)}$,
 is conformally related
 with the metric in Einstein's theory, $ds^2_{\rm (E)}$,
\begin{equation}
    ds^2_{\rm (Q)} = (1+\chi) ds^2_{\rm (E)} , \qquad |\chi/2| \ll 1,
\label{MetricRelationship}\end{equation}
 with $\chi$ satisfying the massive Klein-Gordon equation
 (\ref{chiFieldEq})  in the $ds^2_{\rm (E)}$ metric.
 A consequence of Eq.\ (\ref{MetricRelationship}) is that there will be a
 modification of the  ``Newtonian'' potential equal to $\chi/2$. We will
 have below the oportunity to study its nature and its range in the case
 of topological defects, be them localized or extended sources.

 In general we shall restrict our attention to sufficiently large
 distances, $d$, away from the core, ($d\gg \delta$), but we will keep
 in mind that a proper treatment at short distances requires a proper
 model for the core of the defect itself. In this way we will be able to
 use the existing results in General Relativity for the gravitational
 field  of cosmic strings, monopoles and domain walls which were obtained
 by making use of the above approximations in model Lagrangians with
 symmetry breaking potential of the form (\ref{MexicanHatPot}).

%%---------------------

\subsection{Global monopoles}

 The stress-energy-momentum tensor of a global monopole configuration,
 in regions far away from the core, can be approximated
 by \cite{Barriola,Harari}
\begin{equation} T^t_t= T^r_r \approx - {\eta^2\over r^2},  \qquad
          T^\theta_\theta=T^\varphi_\varphi=0 ,
\label{GlMonopoleSEM}\end{equation}
 while the respective metric in Einstein's theory of gravity is
 (approximately) given by \cite{Barriola,Harari}
\begin{eqnarray}
 ds_{\rm (E)}^2 =-(1&-&\Delta)dt^2 + (1-\Delta)^{-1}dr^2 +d\Omega^2  \cr
     d\Omega^2 &:=&r^2(d\theta^2+\sin^2\theta d\varphi^2 ), \cr
     \Delta &:=&8\pi G\eta^2=\kappa\eta^2.
\label{GlMonopoleEinsteinMetric}\end{eqnarray}
 This metric corresponds to a spacetime with a solid deficit angle:
 test particles are deflected by an angle $\pi\Delta/2$  irrespective of
 their velocity and their impact parameter.
 Here it should be added that a more careful treatment \cite{Harari}
 that takes into account the actual behaviour of the field at the
 monopole core, shows that the metric (\ref{GlMonopoleEinsteinMetric}) gets
 modified by terms which at distances $r\gg \delta=
 (\sqrt{\lambda}\eta)^{-1}$ correspond effectively to a negative
 mass term $M_{\rm eff}$, that is e.g. $g_{tt}\approx (1-\Delta -
 2GM_{\rm eff}/r)$. According to numerical analysis \cite{Harari}
 $M_{\rm eff} \approx -6\pi \sqrt{\lambda}\eta$. Thus, besides the
 topological deflection caused by the solid deficit angle, test
 particles experience also a repulsive radial force $- GM_{\rm eff}/r^2$
 away from the monopole.

 The metric in the quadratic theory is given by
 Eq.\ (\ref{MetricRelationship}) with $\chi$ satisfying
 Eq.\ (\ref{chiFieldEq}).
 Looking for spherically symmetric solutions we find that this equation
 for  $\chi=\chi(r)$ reads
\begin{eqnarray}
 \left\lbrace {1\over r^2}{d\over dr}\left [r^2{d\over dr}\right ] -
 \widehat{m}^2 \right\rbrace \chi(r)&=& {2\Delta \over 3(1-\Delta) r^2},\cr
 \widehat{m}^2&:=&m_0^2/ (1-\Delta).
\label{chiGlMonEq}\end{eqnarray}
 Making use of the Green function for this equation,
\begin{eqnarray}
 G(r,r^\prime)=-{1\over \widehat{m} r r^\prime}
 [&e^{-\widehat{m}r^\prime}&\sinh(\widehat{m}r)\Theta(r^\prime-r) \cr
 + &e^{-\widehat{m}r}&\sinh(\widehat{m}r^\prime)\Theta(r-r^\prime) ],
\label{GlMonGreenFun}\end{eqnarray}
 where the step function $\Theta(z):=\{0,1,1/2\}$ for
 $\{z<0, z>0, z=0\}$ respectively,
 we can write down the solution for $\chi(r)$ in terms of the
 Exponential-Integral $({\rm Ei})$ and Hyperbolic-Sine-Integral
 $({\rm shi})$ functions \cite{Gradshteyn} as
\FL
\begin{equation}
 \chi(r)= {2\Delta\over 3(1-\Delta)}{1\over \widehat{m}r}
 [{\rm Ei}(-\widehat{m}r)\sinh(\widehat{m}r)-e^{-\widehat{m}r}
 {\rm shi}(\widehat{m}r) ] .
\label{chiGlMon}\end{equation}
 Checking numerically the behavior of this function we find that its
 contribution to the `Newtonian' potential $\chi/2$ is an attractive
 one.  In particular, using the asymptotic behavior of the Ei and  shi
 functions \cite{Gradshteyn} we find that at large radial distances
 $r\to\infty$
\begin{equation}
 \chi(r) \approx -{2\over 3} {\Delta\over(m_0r)^2} ,
\label{chiGlMonAs}\end{equation}
 which implies a long range potential, exerting on test particles an
 attractive force $-(2\Delta/3m_0^2)r^{-3}$. Comparing this force to the
 repulsive force due to the core of the monopole we see that the former
 falls off faster by one power of $r$ and thus is negligible at very
 large distances. It overcomes, however, the effect of the latter at a
 distance $r\approx m_0^{-2}/(\lambda\delta)$ and, thus, it can be the
 dominant force within the region
 $\delta\ll r \ll m_0^{-2}/(\lambda\delta)$ which will exist provided
 that $m_0^{-1}\gg \delta$.

 Finally, let us note that the expression (\ref{chiGlMon}) diverges as
 $r\to 0$. This is due to the form of the energy-momentum-tensor in
 Eq.\ (\ref{GlMonopoleSEM}) which is not valid
 at distances comparable to the core of the monopole.

\subsection{Gauge Monopoles}

 A gauge monopole is a spherically symmetric configuration with mass $M$
 and a magnetic charge $g$. Its stress energy momentum tensor can be
 approximated by
\begin{eqnarray}
 T^t_t &=& -{M\over 4\pi} {\delta(r) \over r^2}-{(g/4\pi)^2\over r^4},
 \cr
 T^r_r&=&-T^\theta_\theta=-T^\varphi_\varphi=-{(g/4\pi)^2\over r^4} .
\label{GaugeMonSEM}\end{eqnarray}
 We consider the case where the metric outside the core of the monopole
 matches to a Reissner-Nordstrom one, (see Ref. \cite{Gibbons} for a recent
 review and new results on the gravitational field of monopoles)
\begin{eqnarray}
 ds^2=&-&\left (1-{2GM\over r}+ {G g^2\over 4\pi r^2}\right ) dt^2 \cr
        &+& \left (1-{2G M\over r}+ {G g^2\over 4\pi r^2}\right )^{-1} dr^2
	+ d\Omega^2 .
\label{GaugeMonEinsteinMetric}\end{eqnarray}

 Since the source for the $\chi$ field is the trace of the
 stress-energy-momentum tensor, only the mass term in
 Eq.\ (\ref{GaugeMonSEM}) will contribute.
 Furhermore, if we consider distances sufficiently far from the monopole
 $r\gg \delta \gg GM$, the equation for $\chi$ approximately reads
\begin{equation}
 \left\lbrace {1\over r^2}{d\over dr}\left [r^2{d\over dr}\right ] -
  m_0^2 \right\rbrace \chi(r)= {\kappa M\over 12\pi}
  {\delta(r) \over r^2}.
\label{chiGaugeMonEq}\end{equation}
 Demanding finiteness at radial infinity, this equation has as solution
 the Yukawa fuction
\begin{equation}
 \chi(r)= -{\kappa M\over 12\pi} {e^{-m_0 r}\over r}.
\label{chiGaugeMon}\end{equation}
 Notice that the Newtonian potential of the monopole will be  modified
 by the ammount $\chi/2$ corresponding to an attractive potential
 exponentially decreasing with an $e$-folding term characteristic of a
 massive scalar field with mass $m_0$.

 It worths remarking that the short range
  corrections of Eq.\ (\ref{chiGaugeMon})
 apply also to the external metric of  spherically symmetric mass
 distributions \cite{Teyssandier} such as neutron strars,  giving thus
 rise to ``fifth force'' terms. However, when one deals with black
 holes, the no-hair theorem for $R+R^2$ theories \cite{Whitt} implies
 that corrections of the type (\ref{chiGaugeMon}) are absent.

\subsection{ Global Cosmic Strings}

 As we explained in the introductory part of this section global cosmic
 strings are extended line configurations. The stress-energy-momentum
 tensor for a straight, static, cylindrically symmetric global string
 lying along the $z$-axis is approximately given for $r\gg \delta$ by
\begin{equation}
   T_t^t=T_z^z=T_r^r=-T_\theta^\theta \approx -{\eta^2 \over 2r^2} .
\label{GlStrSEM}\end{equation}
 The respective exact solution for the metric in Einstein's theory has
 been found in  \cite{Cohen}. However, it is quite complicated for the
 purpose of solving the equation (\ref{chiFieldEq}) for the field $\chi$.
 Furthermore, besides this technical problem, the spacetime of a global
 string has true spacetime singularities \cite{Harari88,Gregory}, a fact
 that demands carefull checking of the range of validity of the
 approximation ($\alpha |R| \ll 1$) on which our treatment is based.
 Instead, we prefer to work here in the weak field limit where the
 equation for the $\chi$ field is in Minkowski background metric.

 In the weak field limit of general relativity the metric of the global
 string reads \cite{Harari88}
\begin{eqnarray}
   ds^2_{\rm (E)} = \biggl [1-4G\mu \ln({r\over \delta})
   \biggr ] (-dt^2+dz^2)
      + &dr^2& \cr
      + r^2 \biggl [1-8G\mu\ln({r\over \delta}+c) \biggr ]&d\theta^2&,
\label{GlStringEinsteinMetric}\end{eqnarray}
 Here $\mu:=\pi\eta^2$, $\delta$ is the core width and $c$ is a constant
 of order unity that may partially take into account a global effect of
 the string core.
 Studying the motion of test particles it is seen that the static global
 string exerts a repulsive force $2G\mu/r$ \cite{Harari88}. It is
 interesting to explore how this force is modified in the quadratic
 theory that we are currently considering.

 The equation that $\chi$ statisfies in the weak field limit is
\begin{equation}
 \left\lbrace {1\over r}{d\over dr}\left [r{d\over dr}\right ] -
     m_0^2\right \rbrace \chi(r) = {\kappa\mu\over 3\pi r^2}.
\label{chiGlStringEq}\end{equation}
 The Green function for the homogeneous part of this differential
 equation, with the boundary conditions of finitness at the origin and
 at infinity, is easily found to be
\begin{eqnarray}
 G(r,r^\prime)= &-& K_0(m_0r)I_0(m_0r^\prime)\Theta(r-r^\prime) \cr
                &-& I_0(m_0r)K_0(m_0r^\prime)\Theta(r^\prime -r)
\label{GreenFunction}\end{eqnarray}
 where $\Theta$ is the step function.
 Thus the solution of Eq.\ (\ref{chiGlStringEq}) can be written as
\begin{eqnarray}
    \chi(r)=-{\kappa\mu\over 3\pi} \Bigl [&K_0(m_0r)&\int_\delta^r
      I_0(m_0r^\prime){dr^\prime\over r^\prime}  \cr
      + &I_0(m_0r)&\int_r^\infty
       K_0(m_0r^\prime){dr^\prime\over r^\prime} \Bigr] ,
\label{chiGlString}\end{eqnarray}
 where we have introduced a lower cutoff at $r=\delta$ to cope
 effectively with the divergence that  appears in the first integral if
 we let  $r\to 0$. This divergence is only due to the approximate form
 of the stress-energy-momentum tensor which as we have already stressed
 is not valid near the core of the string.

 The leading term in an asymptotic expansion of Eq.\ (\ref{chiGlString}) at
 large radial distances is
\begin{equation}
 \chi(r) \approx -{\kappa\mu\over 3\pi m_0^2 r^2}, \qquad r\to\infty .
\label{chiGlStringLargeAppr}\end{equation}
 From this expression we conclude that the additional `Newtonian'
 potential in the quadratic theory implies at large distances an
 attractive force $-(\kappa\mu/3m_0^2)r^{-3}$. Due to the slower fall
 off of the original repulsive force, the total force on test particles
 remains repulsive at large distances from the string. At around $r\sim
 m_0^{-1}$ the total force is expected to change sign.

%%---------------------------------

\subsection{Gauge Cosmic Strings}

 Gauge cosmic strings are, in contrast to global ones, localized line
 configurations. The stress-energy-momentum tensor for a static,
 straight along the $z$-axis, gauge cosmic string with line energy
 density $\mu$ is
\FL
\begin{equation}
      T_t^t=T_z^z= -{\mu\over 2\pi(1- \kappa\mu/2)} {\delta(r)\over r},
      \quad T_r^r=T_\theta^\theta=0
\label{GaugeStringSEM}\end{equation}
 with corresponding metric in Einstein's theory \cite{Vilenkin85}
\FL
\begin{equation}
     ds^2_{\rm (E)}= -dt^2+ dz^2 + dr^2 + (1- \kappa\mu/2)^2r^2d\theta^2.
\label{GaugeStringEinsteinMetric}\end{equation}
 Here the polar coordinates $r,\theta$ have the usual range.
 This spacetime is everywhere flat except along the $z$-axis where the
 string is located.
 As one goes around the string one notices an  angle deficit. This
 topological property has  the consequence that test particles which
 localy do not feel any gravitational forces  are, however, deflected by
 the string.

 Let us now turn our attention to  the field $\chi$. It satisfies the
 equation
\FL
\begin{equation}
 \left\lbrace {1\over r}{d\over dr}\left [r{d\over dr}\right ] -
       m_0^2\right \rbrace \chi(r) = {\kappa\mu\over 3\pi
       (1- \kappa\mu/2)}{\delta(r)\over r}.
\label{chiGaugeStringEq}\end{equation}
 which can be easily solved by demanding for the field $\chi$ finiteness
 at infinity and  correct discontinuity at the origin. The solution
 reads
\begin{equation}
     \chi(r)= -{\kappa\mu\over 3\pi(1- \kappa\mu/2)} K_0(m_0 r),
\label{chiGaugeString}\end{equation}
 which can be  easily checked that satisfies  Eq.\ (\ref{chiGaugeStringEq})
 using the small argument asymptotic behavior of the modified Bessel
 function $K_0(z) \approx -\ln(z/2)$.   Finally notice that the field
 $\chi$ decays exponentially fast since at large distances
 $K_0(m_0r)\approx (\pi/2m_0r)^{1/2}\exp(-m_0r)$.
 Very close to the string the expression (\ref{chiGaugeString}) diverges
 logarithmically. Again, as the physical cosmic string has a finite core
 this divergence should not appear in a more proper  treatment near the
 core.

 From Eq.\ (\ref{chiGaugeString}) follows that the `Newtonian' potential
 of the cosmic string spacetime in an attractive, short range one. The
 respective force that the string will exert on test particles is
 $-[\kappa\mu m_0 / 6\pi(1- \kappa\mu/2)] K_1(m_0 r)$. Because of the
 large distance exponential fall off behavior of $K_1(m_0r)\propto
 (m_0r)^{-1/2}\exp(-m_0r)$ it follows that this force is significant
 only close to the string up to  distances $r\sim m_0^{-1}$.

 In closing this subsection let us remark that the result obtained here
 is in agreement with the recent result of Linet and
 Teyssandier \cite{Linet} in the weak field limit where
 $\kappa\mu\ll 1$. These authors have also obtained the cosmic string
 metric in the weak field limit of the  quadratic theory
 (\ref{FullLagrangian}) and (\ref{QuadraticLagrangian}) which
 contains also the massive tensorial field $\psi_{\mu\nu}$ of
 Eqs.\ (\ref{LinearizedEqsa})-(\ref{LinearizedEqsc}).
 In particular they find that the effect of this field on the Newtonian
 potential is a repulsive one with a range set by the inverse mass $m_1$
 of the field $\psi_{\mu\nu}$.

\subsection{Domain Walls}

 The energy content and the gravitational field of domain wall
 configurations in Einstein's theory  has been studied  extensively in
 the literature, see  Ref. \cite{Wang} and references therein.
 We will consider here Vilenkin's vacuum plain domain wall solution
 discussed in   \cite{Vilenkin83}.
 For a domain wall with surface energy density $\sigma$, lying on the
 $|z|=0$ plane,  the stress-energy-momentum tensor is
\begin{equation}
 T^t_t=T^x_x=T^y_y=-\sigma \delta(z),\qquad T^z_z=0,
\label{DWSEM}\end{equation}
 while the respective domain wall spacetime is described by the
 metric \cite{Vilenkin83}
\FL
\begin{eqnarray}
 ds^2_{\rm (E)}=&&(1-\nu|z|)^2\bigl [-dt^2 + e^{2\nu t}(dx^2 + dy^2)
            \bigl ]+ dz^2 , \cr
 &&\nu:=2\pi G\sigma=\kappa\sigma/4.
\label{DWEinsteinMetric}\end{eqnarray}
 Note that some of the metric components are time dependent (no static
 solutions can be found). Test particles in this spacetime are repelled
 with a proper acceleration $\nu$ away from the wall \cite{Vilenkin83},
 a property that we may deduce just by looking at the `Newtonian'
 potential term in the $g_{tt}$ component of the metric
 (\ref{DWEinsteinMetric}).
 Finally we should mention  that at $|z|=\nu^{-1}$ an event horizon
 appears \cite{Vilenkin83}.
 In what follows we restrict our attention to spacetime regions with
 $|z|\leq \nu^{-1}$.

 Although the metric in (\ref{DWEinsteinMetric}) is time dependent we can
 find, however, static solutions to the equation (\ref{chiFieldEq})
 for the field $\chi$ depending only on $|z|$. For such solutions, equation
 (\ref{chiFieldEq})  reduces to the ordinary differential equation
\FL
\begin{equation}
    \left\lbrace (1-\nu|z|)^{-3}{d\over dz}\left
       [(1-\nu|z|)^3{d\over dz}\right]
            - m_0^2 \right\rbrace \chi(z) = 4\nu \delta(z).
\label{chiDWEq}\end{equation}
 The homogeneous part of this equation can be easily transformed into a
 Bessel differential equation for $\widehat{\chi}$ where
 $\chi(|z|):=\widehat{\chi}(\widehat{z})/\widehat{z}$ using the new
 variable $\widehat{z}=\nu^{-1} -|z|$.
 In this way we find that the solution for $\chi(z)$ is, in regions with
 $z\neq 0$, a linear combination of the terms
 $I_1(m_0\widehat{z})/\widehat{z}$ and
 $K_1(m_0\widehat{z})/\widehat{z}$ where $K_1, I_1$ denote modified
 Bessel functions. The coefficients of this solution are determined by
 demanding finiteness at the horizon $\widehat{z}=0$, while on the
 domain wall, $z=0$, the field $\chi$ should be continous  and  have the
 appropriate discontinuity in its first derivative  which, according to
 Eq.\ (\ref{chiDWEq}), is $[{d\over dz}\chi]_{z=0}=4\nu$. Thus we finally
 obtain

\begin{equation}
    \chi(z)= -2\left[{m_0\over\nu} I_2({m_0\over \nu})\right ]^{-1}
    {I_1({m_0\over \nu}[1-\nu|z|]) \over 1-\nu|z|} .
\label{chiDW}\end{equation}

 It is easy to check that this implies  an attractive and  short range
 contribution to the Newtonian potential. This cannot overwhelm the
 original repulsive potential of the domain wall except very close to
 the wall  for $|z|\alt m_0^{-1}$. This should be obvious in writing
 Eq.\ (\ref{chiDW}) in the sensible
 approximation $m_0/\nu \gg 1$ and near the
 domain wall where $\chi$ takes its  largest values
\begin{equation}
     \chi(z)\approx -{2\nu\over m_0}\exp(-m_0|z|) .
\label{chiDWApprox}\end{equation}
 Away from the wall the field $|\chi|$ decreases exponentially
 and  at the horizon
 $\chi$ attains the small value $-1/I_2({m_0\over \nu})$.

\section{Conclusions and remarks}

 We have dealt in this paper mainly with the higher derivative  theory
 (\ref {FullLagrangian}), (\ref{ScalarQuadraticLagrangian})
 with $F=R+\alpha R^2$ in the approximation $\alpha |R|\ll 1$.
 We showed that one can simplify the problem of solving the
 corresponding fourth order field equations using the conformal
 transformation (\ref{ConformalMetric}) which leaves us with  the system
 of field equations (\ref{TransformedFieldEqs}), (\ref{psiFieldEq})
 having only  second order derivatives.
 This  is formally a system of Einstein type  equations
 plus the field equations for a massive scalar field  with mass
 $m_0^2=1/(6\alpha)$ interacting non minimally with gravity.
 We then found that in the approximation $\alpha |R|\ll 1$ the
 gravitational fields in the $R+\alpha R^2$ theory and in the Einstein
 theory are conformally related according to the
 Eqs.\ (\ref{QuadraticFromEinsteinMetric}), (\ref{chiFieldEq}).
 Using this result we looked in Sec. III for solutions representing the
 gravitational field of monopoles, cosmic strings and domain walls

 For localized topological defect configurations as gauge strings, gauge
 monopoles and domain walls we have found  {\sl short range}, {\sl
 attractive} corrections. Their range  $\sim 1/m_0$ is characteristic of
 the pressence of the  massive field.

 For extended sources as global monopoles and global strings we have
 again found {\sl attractive} corrections but with a {\sl long range}.
 Their  fall off rate  depends on the corresponding
 stress-energy-momentum of these defect configurations. In particular,
 for distances $r\gg m_0^{-1}$ is found that the attractive correction
 to the Newtonian potential is $\approx \kappa T/( 6m_0^2)$ where $T$ is
 the trace of the corresponding stress-energy-momentum tensor.

 A more detailed investigation of the gravitational effects of
 topological defects in more general higher order derivative theories is
 in progress and we hope to present the results elsewhere.
 We can however already here reestablish the above given conclusions in
 the $R+\alpha R^2$ theory and at the same time extend them to the case
 of the theory (\ref{FullLagrangian}),(\ref{QuadraticLagrangian}) which,
 for $\beta\neq 0$, contains also a massive tensorial field.
 This will be done by making use of the linearized field equations
 (\ref{LinearizedEqsa}), (\ref{LinearizedEqsb}).
 First  observe that the Newtonian potential $\Phi_N$ will have, besides
 the Einstein term $\Phi_N^{(E)}$, also the contributions
 $\Phi_N^{(\chi)}$ and $\Phi_N^{(\psi_{\mu\nu})}$ from the scalar field
 $\chi$ and massive field $\psi_{\mu\nu}$ respectively
\FL
\begin{equation}
 \Phi_N = \Phi_N^{(E)} + \Phi_N^{(\chi)} + \Phi_N^{(\psi_{\mu\nu})}=
      {1\over 2}[-h^{(E)}_{00} + \chi - \psi_{00}] .
\label{NewtonianPotential}\end{equation}
 For static spacetimes we have from Eqs.\ (\ref{LinearizedEqsa}) and
 (\ref{LinearizedEqsb}) that
\begin{eqnarray}
    \nabla^2 \Phi_N^{(E)} &=& {\kappa\over 2}(\rho + P_1 + P_2 + P_3),
    \cr
   (\nabla^2- m_0^2)\Phi_N^{(\chi)}&=&{\kappa\over 6}(\rho -P_1 -P_2 -P_3),
    \cr
   (\nabla^2- m_1^2)\Phi_N^{(\psi_{\mu\nu})}&=&-{\kappa\over 3}
                                   (2\rho + P_1 + P_2 + P_3) \cr
    m_0^{-2}:=6\alpha +2\beta, && \quad m_1^{-2}:=-\beta,
\label{NewtonianPotentialEqs}\end{eqnarray}
 where $\rho$ denotes the mass density and $P_1,P_2,P_3$ the principal
 pressures of the matter.

{\sl Sign of forces:}
 Look at the r.h.s of these equations. For the topological defect
 configurations  discussed in the present paper $\rho + \sum_iP_i$ is
 $>0$ for gauge monopoles, $0$ for gauge strings and global monopoles,
 and $<0$ for global strings and domain walls. On the other hand
 $\rho -\sum_iP_i > 0$, while $-(2\rho + \sum_iP_i) \leq 0$ with equality
 holding for domain walls. Thus the Einstein contribution in
 Eq.\ (\ref{NewtonianPotential}) is attractive for gauge monopoles,
 zero for gauge strings and global monopoles,
 and repulsive for global strings and
 domain walls. The $\chi$-contribution is always attractive, while the
 $\psi_{\mu\nu}$-contribution is in general repulsive except for domain
 walls where it is  zero.

{\sl Range of forces:}
 From Eqs.\ (\ref {NewtonianPotentialEqs})
 it is clear that the Einstein term
 provides in general a long range interaction due to an effective mass
 density $\rho + \sum_iP_i$. It gives, however, a zero effect for gauge
 cosmic strings and global monopoles.
 For the  contributions of $\chi$ and $\psi_{\mu\nu}$ terms we have
 that:
 \begin{itemize}
\item[(i)] If the stress energy momentum tensor vanishes (or falls off
    sufficiently  rapidly) outside a localized source then at distances
    $d$ from the source, these contributions are of short range $\propto
    \exp(-md) /d^p$ where $m$ stands for the mass $m_0$ (or $m_1$) and
    $p$ is a parameter depending on the symmetry of the spacetime and is
    equal to $0, {1\over 2}$ and $1$ for plain domain walls, strings and
    monopoles respectively.
\item[(ii)] If the sources are not localized then for the $\chi$,
    $\psi_{\mu\nu}$ contributions there are two characteristic regimes:

(a) At distances $d\gg {\rm max}(m_0^{-1}, m_1^{-1})$ the mass terms
          dominate over the derivative terms in the two last equations of
       (\ref{NewtonianPotentialEqs}). Thus, asymptotically at large
       distances we have long range contributions
           $\Phi_{N}^{(\chi)}\approx -\kappa (\rho -\sum_iP_i)/(6m_0^2)=
           \kappa T/(6m_0^2)$, as we found in this paper, and
           $\Phi_{N}^{(\psi_{\mu\nu})}\approx
           \kappa (2\rho + \sum_iP_i)/(3m_1^2)$.

(b) At distances $d\ll{\rm min}(m_0^{-1}, m_1^{-1})$
               the derivative terms dominate over the mass terms.
               The interesting thing to note here is
               that at such distances the total `Newtonian' potential
               in Eq.\ (\ref{NewtonianPotential}) satisfies
               $\nabla^2 \Phi_N \approx 0$. This has implications for the
               differentiability of the spacetime metric and
               implies drastic changes in the
               singularity  structure of gravity at short distances. For
               example, the gravitational potential of a point massive
	       particle is finite at the origin in contrast to the $1/r$
	       Coulomb behavior in the Newtonian theory.
 \end{itemize}

 These considerations are in agreement with the results of the previous
 sections and the results of  \cite{Linet}  for gauge cosmic strings. They
 may be particularly relevant to the study of the evolution of
 topological defects in the very early universe: (a) for structure
 formation scenarios based on global defects where the long range
 modifications of the quadratic theories may play an important role; (b)
 for collisions of cosmic strings where the drastic short range
 modifications may change significantly the predictions of these
 simulations for the evolution parameters of a string network.  Thus it
 is interesting to study further topological defects and  collisions of
 cosmic strings in quadratic gravitational theories and implement
 appropriate modifications in future numerical simulations. The outcome
 of such an investigation confronted with observation, may, among other
 things, allow one to put constraints on the $m_0, m_1$ parameters of
 quadratic gravitational theories.

\acknowledgments
 This work was supported by the European Community DG XII. C.O.L was
 also supported by the Alexander von Humboldt foundation.

\end{document}